\documentclass[preprintnumbers]{revtex4}
\usepackage{amssymb}
\usepackage{amsmath}
\usepackage{graphicx}
\usepackage{calligra}
\usepackage{mathrsfs}
\usepackage{dcolumn}
\usepackage{bm}
\usepackage{subfigure}
\usepackage{color}
\usepackage{CJKutf8}
\numberwithin{equation}{section}
\numberwithin{figure}{section}

\begin{document}
\title{Universal relations with the non-extensive entropy in the RPS and CFT framework}
\author{Hai-Long Zhen$^{1,2}$, Jian-Hua Shi$^{1,2}$, Huai-Fan Li$^{2,3}$, Meng-Sen Ma$^{1,2}$, Yu-Bo Ma$^{1,2}$}
\thanks{\emph{e-mail:} \texttt{dtdx\_yubo.ma@sxdtdx.edu.cn} (corresponding author)}
\affiliation{$^1$Department of Physics, Shanxi Datong University, Datong 037009, China\\
$^2$Institute of Theoretical Physics, Shanxi Datong University, Datong, 037009, China\\
$^3$College of General Education, Shanxi College of Technology, Shuozhou 036000, China\\}

\begin{abstract}

In this paper, an in-depth investigation of the thermodynamic relations for AdS Reissner-Nordstr$\ddot{o}$m black holes immersed in perfect fluid dark matter (PFDM) is presented. This investigation is conducted within both the restricted phase space (RPS) and the conformal field theory (CFT) frameworks. The influence of non-extensive forms of entropy is also systematically examined. Our results demonstrate that the Goon-Penco (GP) relation for charged AdS black holes holds within both the RPS and CFT frameworks, irrespective of the specific form of entropy modification. The conclusion of the present study demonstrates that the GP relation constitutes a universal thermodynamic relation.

\end{abstract}

\maketitle

\section{Introduction}\label{one}

The formulation of black hole thermodynamics has been instrumental in establishing a crucial link between general relativity, classical thermodynamics, and quantum mechanics. The establishment of the four fundamental laws of black hole thermodynamics has positioned this field among the most active frontiers in theoretical physics research \cite{12050559,13015926,12086251,13066233,190610840,240909333,240809500,210906425,150201428,250701010,250623736,250619941}. A comprehensive understanding of black hole thermodynamics holds the potential to clarify the fundamental nature of black holes and reveal characteristics of their internal microstructure. Despite extensive research, critical questions-including the black hole information paradox-remain unresolved.

Recent theoretical advances have introduced the central charge and its chemical potential as a new pair of conjugate thermodynamic variables, significantly enriching the thermodynamic description of black hole. In the dual theory, the central charge is associated with the square of the number of colors \cite{210104145,210502223,024058,220503648}. Within the AdS/CFT correspondence, a novel thermodynamic framework termed restricted phase space thermodynamics (RPST) has been developed and applied to study the homogeneity of the Smarr relation, diverse thermodynamic processes, and phase transitions in various black hole spacetimes \cite{82112,211202386,240415981,230300604,240308991,202585536,230208163,230500267,240215913}. This framework facilitates the exploration of black hole properties and phase transitions in AdS space via holographic duality, providing insights into phenomena such as the Hawking-Page transition, which corresponds to the confinement/deconfinement transition in conformal field theory \cite{250621947,250420486,240702997,211214848,201515077}.

In the domain of quantum gravity, the Weak Gravity Conjecture (WGC), proposed by Vafa, has been advanced as a fundamental principle \cite{0509212}. This conjecture asserts that in any system with mass and charge, gravity must always be the weakest force, expressed mathematically as $M\leq{Q}$. However, extremal black holes satisfy $M=Q$ \cite{0601001}. To uphold the WGC under extreme conditions, it is customary to introduce higher-order derivative terms with specific coefficients. By adjusting these coefficients to meet theoretical requirements, the so-called "Black Hole Weak Gravity Conjecture" is formulated \cite{07120743}. Even when considering the generalized WGC with higher-order derivative corrections, the conjecture continues to hold during black hole evaporation through the emission of neutral particles \cite{10115120,220108380}. Moreover, the correction terms stemming from the weak gravity conjecture can be employed to rule out the existence of naked singularities. That is, the elimination of naked singularities requires that the weak gravity inequality be satisfied \cite{180108546}. To further explore this conjecture, Goon and Penco investigated the thermodynamic relationship between black hole entropy and extremality in a perturbative context \cite{190905254}. This relation connects corrections in free energy to changes in mass, temperature, and entropy, and has since been verified for various types of black holes \cite{20258524,200306785,250719800,035106,231217014,240721329,241102427,250721663}.

\begin{align}\label{1.1}
\frac{\partial {{M}_{ext}}(Q,\eta )}{\partial \eta }&=-\underset{M\to {{M}_{ext}}}{\mathop{\lim }}\,T{{\left( \frac{\partial S(M,Q,\eta )}{\partial \eta } \right)}_{Q,M}},
\end{align}
where $Q$ and $\eta$ represent thermodynamic quantities such as charge, angular momentum, or other parameters, with {\color{red}$\eta$} serving as the perturbative parameter. Here, $M_{ext}$ denotes the extremal mass, and $M$ is the mass of the black hole. Further investigations into universal thermodynamic relations have provided significant insights into the influence of perturbative corrections on black hole properties. These findings enhance the the understanding of fundamental thermodynamic principles and may lead to new discoveries in theoretical physics \cite{100626,201114366,220104071,201105109}.

Meanwhile, a recently developed approach maps the energy function into a two-dimensional space ($r,\theta$), opening a new avenue for studying black hole behavior. This technique enables the investigation of vector field behavior in the mapped space through the computation of winding numbers, yielding interesting insights into geodesic structures and black hole thermodynamics \cite{230616117,241200889,240613784,241104134,250515314,241102875}.

Quantum gravity effects are predicted to induce modifications in the structure of the event horizon, potentially endowing it with a fractal nature and thereby challenging the conventional conception of a smooth horizon. Such modifications lead to corrections in the Bekenstein-Hawking entropy  \cite{241102875}. A natural and actively pursued question is how these entropy corrections affect the GP relation. In this paper, we employ a non-extensive entropy framework to investigate the GP relation within both the RPS and CFT frameworks, and derive a universal expression for this relation.

The paper is organized as follows: In Sec. \ref{two}, the fundamental thermodynamic quantities of RN-PFDM black holes are brief reviewed. In Sec. \ref{three}, the GP relations within both the RPS and CFT frameworks are presented. In Sec. \ref{four}, The GP relations under different corrections to entropy are analyzed. Finally, a concise summary and conclusions are presented in Sec. \ref{five}

\section{The Model} \label{two}

Reissner-Nordstr$\ddot{o}$m AdS black holes enveloped by PFDM represent electrical charged solutions, which are affected by a theoretical type of matter for which there is a theoretical postulation that it constitutes the majority of the universe's mass. These black holes exhibit intriguing thermodynamic behaviors and distinctive phase transition characteristics, which became accessible to through an extended phase space of the system. The action under consideration is that of a gravitational theory which is minimally coupled to a gauge field in the presence of PFDM \cite{240415981,171104538,200903644},

\begin{align}\label{2.1}
f(r)&=1-\frac{2MG}{r}+\frac{{{Q}^{2}}G}{{{r}^{2}}}-\frac{\Lambda }{3}{{r}^{2}}+\frac{\gamma }{r}\ln \left( \frac{r}{\left| \gamma  \right|} \right).
\end{align}

In this analysis, the parameters $M$, $Q$, and $\gamma$ are employed to represent the mass, electric charge, and the influence of (PFDM) on the black hole, respectively. The cosmological constant is given by $\Lambda=-\frac{3}{l^2}$, where $l$ denotes the AdS radius. In the absence of PFDM ($\gamma=0$), the spacetime metric reduces to that of a standard RN AdS black hole.

The black hole under consideration is characterised by the presence of two horizons, designated $r_+$ (the event horizon) and $r_-$ (the inner Cauchy horizon). The mass of the black hole, denoted by  $M$, can be derived by solving $f(r_{\pm})=0$ at the event horizon. After considering the perturbation $\eta$ \cite{20258524,200306785,250719800,035106},

\begin{align}\label{2.2}
f(r)&=1-\frac{2MG}{r}+\frac{{{Q}^{2}}G}{{{r}^{2}}}+\frac{(1+\eta )}{{{l}^{2}}}{{r}^{2}}+\frac{\gamma }{r}\ln \left( \frac{r}{\left| \gamma  \right|} \right),
\end{align}
the energy of the black hole is derived from Eq. (\ref{2.2}),
\begin{align}\label{2.3}
M&=\frac{{{r}_{+}}}{2G}\left[ 1+\frac{{{Q}^{2}}G}{r_{+}^{2}}+\frac{(1+\eta )}{{{l}^{2}}}r_{+}^{2}+\frac{\gamma }{{{r}_{+}}}\ln \left( \frac{{{r}_{+}}}{\left| \gamma  \right|} \right) \right],
\end{align}
the Hawking radiation temperature $T$ and Bekenstein-Hawking entropy $S_{BH}$ are given by
\begin{align}\label{2.4}
T&=\frac{f'({{r}_{+}})}{4\pi }=\frac{1}{4\pi }\left( \frac{1}{{{r}_{+}}}-\frac{{{Q}^{2}}G}{r_{+}^{3}}+\frac{3(1+\eta )}{{{l}^{2}}}{{r}_{+}}+\frac{\gamma }{r_{+}^{2}} \right),~~{{S}_{BH}}=\frac{A}{4G}=\frac{\pi r_{+}^{2}}{G}.
\end{align}

\section{The GP relationship taken Bekenstein-Hawking entropy} \label{three}

\subsection{The GP relationship within RPS framework}\label{threeA}

The first law of thermodynamics for a RN-AdS black hole surrounded by PFDM within the RPS framework is expressed \cite{240415981},
\begin{align}\label{3.1}
dM=Td{{S}_{BH}}+\tilde{\varphi }d\tilde{Q}+\mu dC+\tilde{\prod }d\tilde{\gamma },
\end{align}
where $\tilde{\varphi}$ and $\tilde{Q}$ denote the appropriately re-scaled electric potential and electric charge, respectively, and $\tilde{\prod}$ denotes the conjugate variable to $\tilde{\gamma}$.

In the context of the black hole under consideration, the parameters $M$, $T$ and $S_\text{BH}$ correspond to the mass, Hawking temperature, and Bekenstein-Hawking entropy, respectively. Furthermore, these thermodynamic quantities can generally be described in terms of their dual counterparts in the CFT framework as follows:
\begin{align}\label{3.2}
\tilde{Q}=\frac{Ql}{\sqrt{G}},~~\tilde{\varphi }=\frac{\varphi \sqrt{G}}{l},~~\tilde{\gamma }=\frac{\gamma {{l}^{2}}}{G},~~\tilde{\prod }=\frac{\prod G}{{{l}^{2}}},
\end{align}
some thermodynamic quantities can be described in terms of their dual counterparts in the RPS framework as follows:
\begin{align}\label{3.3}
C=\frac{{{l}^{2}}}{G},~~\tilde{Q}=Q\sqrt{C},~~{{r}_{+}}=l\sqrt{\frac{{{S}_{BH}}}{\pi C}},~~\tilde{\gamma }=\gamma C.
\end{align}

According to Eqs. (\ref{2.3}) and (\ref{3.3}), the following thermodynamic quantities are obtained:
\begin{align}\label{3.4}
M&=\frac{\pi C{{S}_{BH}}+{{\pi }^{2}}{{{\tilde{Q}}}^{2}}+(1+\eta )S_{BH}^{2}}{2{{\pi }^{3/2}}l\sqrt{C{{S}_{BH}}}}+\frac{{\tilde{\gamma }}}{2{{l}^{2}}}\ln \left( \sqrt{\frac{C{{S}_{BH}}}{\pi }}\frac{l}{{\tilde{\gamma }}} \right),\notag \\
T&={{\left( \frac{\partial M}{\partial {{S}_{BH}}} \right)}_{\tilde{Q},C,\tilde{\gamma }}}=\frac{\pi C{{S}_{BH}}-{{\pi }^{2}}{{{\tilde{Q}}}^{2}}+3(1+\eta )S_{BH}^{^{2}}}{4{{\pi }^{3/2}}lS_{BH}^{3/2}\sqrt{C}}+\frac{{\tilde{\gamma }}}{4{{l}^{2}}{{S}_{BH}}},\notag \\
\tilde{\varphi }&={{\left( \frac{\partial M}{\partial \tilde{Q}} \right)}_{{{S}_{BH}},C,\tilde{\gamma }}}=\sqrt{\frac{\pi }{{{S}_{BH}}C}}\frac{{\tilde{Q}}}{l},~~\tilde{\prod }={{\left( \frac{\partial M}{\partial \tilde{\gamma }} \right)}_{\tilde{Q},C,{{S}_{BH}}}}=\frac{1}{2{{l}^{2}}}\left[ -1+\ln \left( \sqrt{\frac{{{S}_{BH}}C}{\pi }}\frac{l}{{\tilde{\gamma }}} \right) \right],\notag \\
\mu& ={{\left( \frac{\partial M}{\partial C} \right)}_{\tilde{Q},{{S}_{BH}},\tilde{\gamma }}}=\frac{\pi C{{S}_{BH}}-{{\pi }^{2}}{{{\tilde{Q}}}^{2}}-(1+\eta )S_{BH}^{2}}{4{{\pi }^{3/2}}lC\sqrt{C{{S}_{BH}}}}+\frac{{\tilde{\gamma }}}{4{{l}^{2}}C},
\end{align}
according to Eq. (\ref{3.4}), the following equation is obtained:
\begin{align}\label{3.5}
\eta =\frac{2l{{\pi }^{3/2}}{{C}^{1/2}}}{S_{BH}^{3/2}}\left[ M-\frac{1}{2l}{{\left( \frac{{{S}_{BH}}C}{\pi } \right)}^{1/2}}-\frac{{{{\tilde{Q}}}^{2}}}{2l}{{\left( \frac{\pi }{{{S}_{BH}}C} \right)}^{1/2}}-\frac{{\tilde{\gamma }}}{2{{l}^{2}}}\ln \left( \frac{l}{{\tilde{\gamma }}}{{\left( \frac{C{{S}_{BH}}}{\pi } \right)}^{1/2}} \right) \right]-1.
\end{align}
In the event that $M=M_0$(being constant), when $\tilde{Q}$, $C$, $\tilde{\gamma}$ and $l$ are held constant, according to Eq. (\ref{3.5}), the following equation is obtained:
\begin{align}\label{3.6}
{{\left( \frac{\partial S_{BH}}{\partial \eta } \right)}_{M}}=-\frac{\frac{3S_{BH}^{1/2}}{2l{{\pi }^{3/2}}{{C}^{1/2}}}}{\left( \frac{1}{4l{{S}_{BH}}}{{\left( \frac{{{S}_{BH}}C}{\pi } \right)}^{1/2}}-\frac{{{{\tilde{Q}}}^{2}}}{4l{{S}_{BH}}}{{\left( \frac{\pi }{{{S}_{BH}}C} \right)}^{1/2}}+\frac{3(1+\eta )S_{BH}^{1/2}}{4l{{\pi }^{3/2}}{{C}^{1/2}}}+\frac{{\tilde{\gamma }}}{4{{l}^{2}}{{S}_{BH}}} \right)}=-\frac{\frac{3S_{BH}^{3/2}}{2l{{\pi }^{3/2}}{{C}^{1/2}}}}{T}.
\end{align}
When $\tilde{Q}$, $C$, $\tilde{\gamma}$ and $l$ are held constant, according to Eq. (\ref{3.4}), the following equation is obtained:
\begin{align}\label{3.7}
\frac{\partial M}{\partial \eta }=\frac{S_{BH}^{3/2}}{2l{{\pi }^{3/2}}{{C}^{1/2}}}.
\end{align}
From Eqs. (\ref{3.6}) and (\ref{3.7}), The GP relation within the RPS framework is obtained:
\begin{align}\label{3.8}
\frac{\partial {{M}_{0}}}{\partial \eta }=\frac{S_{BH}^{3/2}}{2l{{\pi }^{3/2}}{{C}^{1/2}}}=\underset{M\to {{M}_{0}}}{\mathop{\lim }}\,-T\frac{\partial {{S}_{BH}}}{\partial \eta }.
\end{align}

\subsection{The GP relationship within CFT framework}

The first law of thermodynamics for a RN-AdS black hole surrounded by PFDM within the CFT framework is expressed \cite{240415981},
\begin{align}\label{3.9}
dE=\hat{T}d{{S}_{BH}}+\hat{\phi }d\hat{Q}-pdv+\mu dC+\hat{\prod }d\hat{\gamma },
\end{align}
where
\begin{align}\label{3.10}
E=\frac{M}{\omega },~~\hat{T}=\frac{T}{\omega },~~\hat{\phi }=\frac{\phi \sqrt{G}}{\omega l},~~\hat{Q}=\frac{Ql}{\sqrt{G}},~~\hat{\prod }=\frac{\prod \sqrt{G}}{\omega l},~~\hat{\gamma }=\frac{\gamma l}{\sqrt{G}},
\end{align}
if $B=\frac{\tilde{\gamma}}{l}$, these thermodynamic quantities within the CFT can be calculated as follows:
\begin{align}\label{3.11}
E&=\frac{4\pi C{{S}_{BH}}+{{\pi }^{2}}{{{\hat{Q}}}^{2}}+(1+\eta )S_{BH}^{2}}{2\pi \sqrt{C{{S}_{BH}}v}}+\frac{2B\sqrt{\pi C}}{\sqrt{v}}\ln \left( \sqrt{\frac{{{S}_{BH}}}{\pi }}\frac{1}{B} \right),\notag \\
\hat{T}&={{\left( \frac{\partial E}{\partial {{S}_{BH}}} \right)}_{\hat{Q},C,B,v}}=\frac{4\pi C{{S}_{BH}}-{{\pi }^{2}}{{{\hat{Q}}}^{2}}+3(1+\eta )S_{BH}^{^{2}}}{4\pi S_{BH}^{3/2}\sqrt{Cv}}+\frac{B\sqrt{\pi C}}{{{S}_{BH}}\sqrt{v}},\notag \\
\hat{\phi }&={{\left( \frac{\partial E}{\partial \hat{Q}} \right)}_{{{S}_{BH}},C,B,v}}=\frac{\pi \hat{Q}}{\sqrt{{{S}_{BH}}Cv}},\notag \\ \hat{\prod }&={{\left( \frac{\partial M}{\partial \hat{\gamma }} \right)}_{\hat{Q},C,{{S}_{BH}},v}}=\frac{B\sqrt{\pi C}}{\hat{\gamma }{{S}_{BH}}\sqrt{v}},~~-p={{\left( \frac{\partial E}{\partial v} \right)}_{\hat{Q},{{S}_{BH}},C,B}}=-\frac{E}{2v},\notag \\
\mu& ={{\left( \frac{\partial E}{\partial C} \right)}_{\tilde{Q},{{S}_{BH}},B,v}}=\frac{4\pi C{{S}_{BH}}-{{\pi }^{2}}{{{\tilde{Q}}}^{2}}-(1+\eta )S_{BH}^{2}}{4\pi C\sqrt{C{{S}_{BH}}v}}+\frac{\sqrt{\pi }B}{\sqrt{Cv}}\ln \left( \frac{\sqrt{{{S}_{BH}}}}{\sqrt{\pi }B} \right).
\end{align}
According to Eq. (\ref{3.11}), the following equation is obtained:
\begin{align}\label{3.12}
\eta& =\frac{2\pi \sqrt{Cv}}{S_{BH}^{3/2}}E-\frac{4\pi C}{{{S}_{BH}}}-\frac{{{\pi }^{2}}{{{\hat{Q}}}^{2}}}{S_{BH}^{2}}-4{{\pi }^{3/2}}C\frac{B}{S_{BH}^{3/2}}\ln \left( \sqrt{\frac{{{S}_{BH}}}{\pi }}\frac{1}{B} \right)-1.
\end{align}
When $\tilde{Q}$, $C$, $B$, $\nu$ and $l$ are held constant, according to Eq. (\ref{3.11}), the following equation is obtained:
\begin{align}\label{3.13}
\frac{\partial E}{\partial \eta }&=\frac{S_{BH}^{3/2}}{2\pi \sqrt{Cv}}.
\end{align}
In the event that $E=E_0$(being constant), when $\tilde{Q}$, $C$, $B$ and $l$ are held constant, according to Eq. (\ref{3.12}), the following equation is obtained:
\begin{align}\label{3.14}
{{\left( \frac{\partial S_{BH}}{\partial \eta } \right)}_{E}}&=-\frac{\frac{S_{BH}^{3/2}}{2\pi \sqrt{Cv}}}{\frac{4\pi C{{S}_{BH}}-{{\pi }^{2}}{{{\hat{Q}}}^{2}}+3(1+\eta )S_{BH}^{^{2}}}{4\pi S_{BH}^{3/2}\sqrt{Cv}}+\frac{B\sqrt{\pi C}}{{{S}_{BH}}\sqrt{v}}}=-\frac{\frac{S_{BH}^{3/2}}{2\pi \sqrt{Cv}}}{{\hat{T}}}.
\end{align}
According to Eqs. (\ref{3.13}) and (\ref{3.14}), The GP relation within the CFT framework is obtained:
\begin{align}\label{3.15}
\frac{\partial {{E}_{0}}}{\partial \eta }&=\frac{S_{BH}^{3/2}}{2\pi \sqrt{Cv}}=\underset{E\to {{E}_{0}}}{\mathop{\lim }}\,-\hat{T}\frac{\partial {{S}_{BH}}}{\partial \eta }.
\end{align}

\section{non-extensive entropy and GP relation}\label{four}

\subsection{Universality relation from Barrow entropy}

Barrow introduced a modification to the concept of black hole entropy by accounting for quantum gravitational effects. These effects have been demonstrated to challenge the conventional picture of a smooth and uniform event horizon. This approach incorporates a fractal parameter, denoted by $Delta$, to quantify the degree of structural distortion induced by quantum gravity on the horizon. The resulting Barrow entropy provides a generalized expression for black hole entropy \cite{241102875,200409444,170309355,241212132,241212137},
\begin{align}\label{4.1}
{{S}_{B}}={{({{S}_{BH}})}^{1+\tfrac{\Delta }{2}}},~~{{S}_{BH}}=S_{B}^{\tfrac{2}{2+\Delta }},
\end{align}
where, $\Delta$ is constrained to the interval $[0,1]$. For $\Delta$=0, the entropy reverts to the standard Bekenstein-Hawking form, thereby indicating the absence of fractal structure.

Substituting Eq. (\ref{4.1}) into (\ref{3.4}), the following thermodynamic quantities are obtained:
\begin{align}\label{4.2}
{{\tilde{M}}_{B}}&=\frac{\sqrt{C}}{2{{\pi }^{1/2}}l}S_{B}^{\frac{1}{2+\Delta }}+\frac{{{\pi }^{1/2}}{{{\tilde{Q}}}^{2}}}{2l\sqrt{C}}S_{B}^{-\frac{1}{2+\Delta }}+\frac{(1+\eta )S_{B}^{\tfrac{3}{2+\Delta }}}{2{{\pi }^{3/2}}l\sqrt{C}}+\frac{{\tilde{\gamma }}}{2{{l}^{2}}}\ln \left( \sqrt{\frac{C}{\pi }}\frac{l}{{\tilde{\gamma }}}S_{B}^{\tfrac{1}{2+\Delta }} \right),\notag \\
{{\tilde{T}}_{B}}&={{\left( \frac{\partial {{{\tilde{M}}}_{B}}}{\partial {{S}_{B}}} \right)}_{\tilde{Q},C,\tilde{\gamma }}}=\left( \frac{2}{2+\Delta } \right)\left( \frac{\pi CS_{B}^{\tfrac{2}{2+\Delta }}-{{\pi }^{2}}{{{\tilde{Q}}}^{2}}+3(1+\eta )S_{B}^{\tfrac{4}{2+\Delta }}}{4{{\pi }^{3/2}}lS_{B}^{\tfrac{5}{2+\Delta }}\sqrt{C}}+\frac{{\tilde{\gamma }}}{4{{l}^{2}}S_{B}^{\tfrac{4}{2+\Delta }}} \right)S_{B}^{-\tfrac{\Delta }{2+\Delta }}.
\end{align}
Accordingly, the first law of thermodynamics (Eq. (\ref{3.1})) that is satisfied by the black hole when the modified entropy is the state parameter of the black hole thermodynamic system, is reformulated as follows:
\begin{align}\label{4.3}
d{{\tilde{M}}_{B}}&={{\tilde{T}}_{B}}d{{S}_{B}}+\tilde{\varphi }d\tilde{Q}+\mu dC+\tilde{\prod }d\tilde{\gamma }.
\end{align}
When $\tilde{Q}$, $C$, $\tilde{\gamma}$ and $l$ are held constant, according to Eq. (\ref{4.2}), the following equation is obtained:
\begin{align}\label{4.4}
\frac{\partial {{{\tilde{M}}}_{B}}}{\partial \eta }&=\frac{S_{B}^{\tfrac{3}{2+\Delta }}}{2{{\pi }^{3/2}}l\sqrt{C}},
\end{align}
and
\begin{align}\label{4.5}
\eta =\frac{2{{\pi }^{3/2}}l\sqrt{C}{{{\tilde{M}}}_{B}}}{S_{B}^{\tfrac{3}{2+\Delta }}}\pi -\frac{C}{S_{B}^{\tfrac{2}{2+\Delta }}}-{{\pi }^{2}}{{\tilde{Q}}^{2}}S_{B}^{-\frac{4}{2+\Delta }}-\frac{{{\pi }^{3/2}}\sqrt{C}}{S_{B}^{\tfrac{3}{2+\Delta }}}\frac{{\tilde{\gamma }}}{l}\ln \left( \sqrt{\frac{C}{\pi }}\frac{l}{{\tilde{\gamma }}}S_{B}^{\tfrac{1}{2+\Delta }} \right)-1.
\end{align}
In the event that $\tilde{M}_B=\tilde{M}_B^0$(being constant), when $\tilde{Q}$, $C$, $\tilde{\gamma}$, and $l$ are held constant, according to Eq. (\ref{4.5}), the following equation is obtained:
\begin{align}\label{4.6}
{{\left( \frac{\partial {{S}_{B}}}{\partial \eta } \right)}_{M_{B}^{0}}}=-\frac{\frac{S_{B}^{\tfrac{3}{2+\Delta }}}{2{{\pi }^{3/2}}l\sqrt{C}}}{\left( \frac{2}{2+\Delta } \right)\left( \frac{\pi CS_{B}^{\tfrac{2}{2+\Delta }}-{{\pi }^{2}}{{{\tilde{Q}}}^{2}}+3(1+\eta )S_{B}^{\tfrac{4}{2+\Delta }}}{4{{\pi }^{3/2}}lS_{B}^{\tfrac{5}{2+\Delta }}\sqrt{C}}+\frac{{\tilde{\gamma }}}{4{{l}^{2}}S_{B}^{\tfrac{4}{2+\Delta }}} \right)S_{B}^{-\tfrac{\Delta }{2+\Delta }}}.
\end{align}
In the event that entropy assumes the Barrow entropy, according to Eqs. (\ref{4.2}), (\ref{4.4}) and (\ref{4.6}), the GP relation within the RPS framework is obtained:
\begin{align}\label{4.7}
\frac{\partial \tilde{M}_{B}^{0}}{\partial \eta }=\frac{S_{B}^{\tfrac{3}{2+\Delta }}}{2{{\pi }^{3/2}}l\sqrt{C}}=\underset{{{{\tilde{M}}}_{B}}\to \tilde{M}_{B}^{0}}{\mathop{\lim }}\,-{{\tilde{T}}_{B}}\left( \frac{\partial {{S}_{B}}}{\partial \eta } \right)
\end{align}

Substituting Eq. (\ref{4.1}) into Eq. (\ref{3.11}), the following equation is obtained:
\begin{align}\label{4.8}
{{\hat{E}}_{B}}&=\frac{4\pi CS_{B}^{\tfrac{2}{2+\Delta }}+{{\pi }^{2}}{{{\hat{Q}}}^{2}}+(1+\eta )S_{B}^{\tfrac{4}{2+\Delta }}}{2\pi S_{B}^{\tfrac{1}{2+\Delta }}\sqrt{Cv}}+\frac{2B\sqrt{\pi C}}{\sqrt{v}}\ln \left( \frac{S_{B}^{\tfrac{1}{2+\Delta }}}{\sqrt{\pi }B} \right),\notag \\
{{\hat{T}}_{B}}&={{\left( \frac{\partial E}{\partial {{S}_{B}}} \right)}_{\hat{Q},C,B,v}}=\left( \frac{2}{2+\Delta } \right)\left( \frac{4\pi CS_{B}^{\tfrac{2}{2+\Delta }}-{{\pi }^{2}}{{{\hat{Q}}}^{2}}+3(1+\eta )S_{B}^{\tfrac{4}{2+\Delta }}}{4\pi S_{B}^{\tfrac{3}{2+\Delta }}\sqrt{Cv}}+\frac{B\sqrt{\pi C}}{S_{B}^{\tfrac{2}{2+\Delta }}\sqrt{v}} \right)S_{B}^{-\tfrac{\Delta }{2+\Delta }}.
\end{align}
Accordingly, the first law of thermodynamics (Eq. (\ref{3.9})) that is satisfied by the black hole when the modified entropy is the state parameter of the black hole thermodynamic system, is reformulated as follows:
\begin{align}\label{4.9}
d{{\hat{E}}_{B}}={{\hat{T}}_{B}}d{{S}_{B}}+\hat{\phi }d\hat{Q}-pdv+\mu dC+\hat{\prod }d\hat{\gamma }.
\end{align}
When $\tilde{Q}$, $C$, $B$, and $\nu$ are held constant, according to Eq.(\ref{4.8}), the following equation is obtained:
\begin{align}\label{4.10}
\frac{\partial {{{\hat{E}}}_{B}}}{\partial \eta }=\frac{S_{B}^{\tfrac{3}{2+\Delta }}}{2\pi \sqrt{Cv}},
\end{align}
\begin{align}\label{4.11}
\eta =2\pi S_{B}^{-\tfrac{3}{2+\Delta }}\sqrt{Cv}{{\hat{E}}_{B}}-4\pi CS_{B}^{-\tfrac{2}{2+\Delta }}-{{\pi }^{2}}{{\hat{Q}}^{2}}S_{B}^{-\tfrac{4}{2+\Delta }}-4{{\pi }^{3/2}}S_{B}^{\tfrac{3}{2+\Delta }}C\ln \left( \frac{S_{B}^{\tfrac{1}{2+\Delta }}}{\sqrt{\pi }B} \right)-1.
\end{align}
In the event that $\hat{E}_B=\hat{E}_B^0$(being constat), when $\tilde{Q}$, $C$, $B$, and $\nu$ are held constant, according to Eq. (\ref{4.11}), the following equation is obtained:
\begin{align}\label{4.12}
{{\left( \frac{\partial {{S}_{B}}}{\partial \eta } \right)}_{\hat{E}_{B}^{0}}}=-\frac{\frac{S_{B}^{\tfrac{3}{2+\Delta }}}{2\pi \sqrt{Cv}}}{\left( \frac{2}{2+\Delta } \right)\left( \frac{4\pi CS_{B}^{\tfrac{2}{2+\Delta }}-{{\pi }^{2}}{{{\hat{Q}}}^{2}}+3(1+\eta )S_{B}^{\tfrac{4}{2+\Delta }}}{4\pi S_{B}^{\tfrac{3}{2+\Delta }}\sqrt{Cv}}+\frac{B\sqrt{\pi C}}{S_{B}^{\tfrac{2}{2+\Delta }}\sqrt{v}} \right)S_{B}^{-\tfrac{\Delta }{2+\Delta }}}=-\frac{\frac{S_{B}^{\tfrac{3}{2+\Delta }}}{2\pi \sqrt{Cv}}}{{{{\hat{T}}}_{B}}}
\end{align}
In the event that entropy assumes the Barrow entropy, according to Eqs. (\ref{4.10}) and (\ref{4.12}), the GP relation within the CFT framework is obtained:
\begin{align}\label{4.13}
\frac{\partial {{{\hat{E}}}_{B0}}}{\partial \eta }=\underset{{{{\hat{E}}}_{B}}\to {{{\hat{E}}}_{B0}}}{\mathop{\lim }}\,-{{\hat{T}}_{B}}\left( \frac{\partial {{S}_{B}}}{\partial \eta } \right)
\end{align}

\subsection{Universality relation from Renyi entropy}

R$\acute{e}$nyi entropy is a type of non-extensive entropy that has been applied to the study of black hole thermodynamics. It is characterized by a parameter that modulates the degree of non-extensiveness, which is required to remain within a specified range to ensure the well-defined behavior of the entropy function. In the context of black holes, R$\acute{e}$nyi entropy provides a generalized framework for analyzing thermodynamic properties that extends beyond the traditional Boltzmann-Gibbs statistics \cite{241102875,221105989,250100955},
\begin{align}\label{4.14}
{{S}_{R}}=\frac{1}{\lambda }\ln (1+\lambda {{S}_{BH}}),~~{{S}_{BH}}=\frac{{{e}^{\lambda {{S}_{R}}}}-1}{\lambda },
\end{align}
the parameter $\lambda$ in non-extensive entropy is essential in shaping the entropy formulation. To ensure the entropy function remains well-defined, $\lambda$ must satisfy ${-\infty}<{\lambda}<1$; outside this interval, the function becomes convex and hence ill-defined. In the context of black hole thermodynamics based on R$\acute{e}$nyi statistics, the entropy $SR$ is well defined when $0 \leq \lambda < 1$. Within this range, $\lambda$ exhibits favorable thermodynamic properties, as demonstrated by recent studies. Notably, as the R$\acute{e}$nyi parameter $\lambda \to 0$, the generalized off-shell free energy reduces to the classical Boltzmann-Gibbs statistics.

Substituting Eq. (\ref{4.14}) into Eq. (\ref{3.4}), the following equation is obtained:
\begin{align}\label{4.15}
{{\tilde{M}}_{R}}&=\frac{\pi C\frac{{{e}^{\lambda {{S}_{R}}}}-1}{\lambda }+{{\pi }^{2}}{{{\tilde{Q}}}^{2}}+(1+\eta ){{\left( \frac{{{e}^{\lambda {{S}_{R}}}}-1}{\lambda } \right)}^{2}}}{2{{\pi }^{3/2}}l\sqrt{C}{{\left( \frac{{{e}^{\lambda {{S}_{R}}}}-1}{\lambda } \right)}^{1/2}}}+\frac{{\tilde{\gamma }}}{2{{l}^{2}}}\ln \left( \sqrt{\frac{C}{\pi }}\frac{l}{{\tilde{\gamma }}}{{\left( \frac{{{e}^{\lambda {{S}_{R}}}}-1}{\lambda } \right)}^{1/2}} \right),\notag \\
{{\tilde{T}}_{R}}&={{\left( \frac{\partial {{{\tilde{M}}}_{R}}}{\partial {{S}_{R}}} \right)}_{\tilde{Q},C,\tilde{\gamma }}}=\frac{\pi C\frac{{{e}^{\lambda {{S}_{R}}}}-1}{\lambda }-{{\pi }^{2}}{{{\tilde{Q}}}^{2}}+3(1+\eta ){{\left( \frac{{{e}^{\lambda {{S}_{R}}}}-1}{\lambda } \right)}^{2}}}{4{{\pi }^{3/2}}l\sqrt{C}}{{\left( \frac{\lambda }{{{e}^{\lambda {{S}_{R}}}}-1} \right)}^{3/2}}{{e}^{\lambda {{S}_{R}}}}+\frac{{\tilde{\gamma }}}{4{{l}^{2}}}\left( \frac{\lambda }{{{e}^{\lambda {{S}_{R}}}}-1} \right){{e}^{\lambda {{S}_{R}}}}.
\end{align}
Accordingly, the first law of thermodynamics (Eq. (\ref{3.1})) that is satisfied by the black hole when the modified entropy is the state parameter of the black hole thermodynamic system, is reformulated as follows:
\begin{align}\label{4.16}
d{{\tilde{M}}_{R}}={{\tilde{T}}_{R}}d{{S}_{R}}+\tilde{\varphi }d\tilde{Q}+\mu dC+\tilde{\prod }d\tilde{\gamma }.
\end{align}
When $\tilde{Q}$, $C$, $\tilde{\gamma}$, and $l$ are held constant, according to Eq.(\ref{4.15}), the following equation is obtained:
\begin{align}\label{4.17}
\frac{\partial {{{\tilde{M}}}_{R}}}{\partial \eta }=\frac{1}{2{{\pi }^{3/2}}l\sqrt{C}}{{\left( \frac{{{e}^{\lambda {{S}_{R}}}}-1}{\lambda } \right)}^{3/2}},
\end{align}
with
\begin{align}\label{4.18}
\eta& =\frac{2{{\pi }^{3/2}}l\sqrt{C}{{\left( \frac{{{e}^{\lambda {{S}_{R}}}}-1}{\lambda } \right)}^{1/2}}{{{\tilde{M}}}_{R}}-\pi C\frac{{{e}^{\lambda {{S}_{R}}}}-1}{\lambda }-{{\pi }^{2}}{{{\tilde{Q}}}^{2}}}{{{\left( \frac{{{e}^{\lambda {{S}_{R}}}}-1}{\lambda } \right)}^{2}}}, \notag \\
&-{{\pi }^{3/2}}\sqrt{C}{{\left( \frac{{{e}^{\lambda {{S}_{R}}}}-1}{\lambda } \right)}^{-3/2}}\frac{{\tilde{\gamma }}}{l}\ln \left( \sqrt{\frac{C}{\pi }}\frac{l}{{\tilde{\gamma }}}{{\left( \frac{{{e}^{\lambda {{S}_{R}}}}-1}{\lambda } \right)}^{1/2}} \right)-1.
\end{align}
In the event that ${{{\tilde{M}}}_{R}}={{{\tilde{M}}}_{R}^{0}}$(being constant), when $\tilde{Q}$, $C$, $\tilde{\gamma}$, and $l$ are held constant, according to Eq. (\ref{4.18}), the following equation is obtained:
\begin{align}\label{4.19}
{{\left( \frac{\partial {{S}_{R}}}{\partial \eta } \right)}_{\tilde{M}_{R}^{0}}}&=-\frac{\frac{1}{2{{\pi }^{3/2}}l\sqrt{C}}{{\left( \frac{{{e}^{\lambda {{S}_{R}}}}-1}{\lambda } \right)}^{3/2}}}{\frac{\pi C\frac{{{e}^{\lambda {{S}_{R}}}}-1}{\lambda }-{{\pi }^{2}}{{{\tilde{Q}}}^{2}}+3(1+\eta ){{\left( \frac{{{e}^{\lambda {{S}_{R}}}}-1}{\lambda } \right)}^{2}}}{4{{\pi }^{3/2}}l\sqrt{C}}{{\left( \frac{\lambda }{{{e}^{\lambda {{S}_{R}}}}-1} \right)}^{3/2}}{{e}^{\lambda {{S}_{R}}}}+\frac{{\tilde{\gamma }}}{4{{l}^{2}}}\left( \frac{\lambda }{{{e}^{\lambda {{S}_{R}}}}-1} \right){{e}^{\lambda {{S}_{R}}}}}.
\end{align}
In the event that entropy assumes the R$\acute{e}$nyi entropy, according to Eqs. (\ref{4.15}), (\ref{4.17}), and (\ref{4.19}), the GP relation within the RPS framework is obtained:
\begin{align}\label{4.20}
\frac{\partial \tilde{M}_{R}^{0}}{\partial \eta }=\frac{1}{2{{\pi }^{3/2}}l\sqrt{C}}{{\left( \frac{{{e}^{\lambda {{S}_{R}}}}-1}{\lambda } \right)}^{3/2}}=\underset{{{{\tilde{M}}}_{R}}\to \tilde{M}_{R}^{0}}{\mathop{\lim }}\,-{{\tilde{T}}_{R}}\left( \frac{\partial {{S}_{R}}}{\partial \eta } \right).
\end{align}

Substituting Eq. (\ref{4.14}) into (\ref{3.11}), the following equation is obtained:
\begin{align}\label{4.21}
{{\hat{E}}_{R}}=\frac{4\pi C\left( \frac{{{e}^{\lambda {{S}_{R}}}}-1}{\lambda } \right)+{{\pi }^{2}}{{{\hat{Q}}}^{2}}+(1+\eta ){{\left( \frac{{{e}^{\lambda {{S}_{R}}}}-1}{\lambda } \right)}^{2}}}{2\pi \sqrt{Cv}{{\left( \frac{{{e}^{\lambda {{S}_{R}}}}-1}{\lambda } \right)}^{1/2}}}+\frac{2B\sqrt{\pi C}}{\sqrt{v}}\ln \left( \frac{1}{\sqrt{\pi }B}{{\left( \frac{{{e}^{\lambda {{S}_{R}}}}-1}{\lambda } \right)}^{1/2}} \right),
\end{align}
\begin{align}\label{4.22}
{{\hat{T}}_{R}}={{\left( \frac{\partial {{{\hat{E}}}_{R}}}{\partial {{S}_{R}}} \right)}_{\hat{Q},C,B,v}}=\frac{4\pi C\left( \frac{{{e}^{\lambda {{S}_{R}}}}-1}{\lambda } \right)-{{\pi }^{2}}{{{\hat{Q}}}^{2}}+3(1+\eta ){{\left( \frac{{{e}^{\lambda {{S}_{R}}}}-1}{\lambda } \right)}^{2}}}{4\pi \sqrt{Cv}{{\left( \frac{{{e}^{\lambda {{S}_{R}}}}-1}{\lambda } \right)}^{3/2}}}{{e}^{\lambda {{S}_{R}}}}+\frac{B\sqrt{\pi C}}{{{S}_{BH}}\sqrt{v}}\left( \frac{\lambda }{{{e}^{\lambda {{S}_{R}}}}-1} \right){{e}^{\lambda {{S}_{R}}}}.
\end{align}
Accordingly, the first law of thermodynamics (Eq. (\ref{3.9})) that is satisfied by the black hole when the modified entropy is the state parameter of the black hole thermodynamic system, is reformulated as follows:
\begin{align}\label{4.23}
d{{\hat{E}}_{R}}={{\hat{T}}_{R}}d{{S}_{R}}+\hat{\phi }d\hat{Q}-pdv+\mu dC+\hat{\prod }d\hat{\gamma }.
\end{align}
When $\tilde{Q}$, $C$, $B$ and $\nu$ are held constant, according to Eq.(\ref{4.21}), the following equation is obtained:
\begin{align}\label{4.24}
\frac{\partial {{{\hat{E}}}_{R}}}{\partial \eta }=\frac{1}{2\pi \sqrt{Cv}}{{\left( \frac{{{e}^{\lambda {{S}_{R}}}}-1}{\lambda } \right)}^{3/2}},
\end{align}
and
\begin{align}\label{4.25}
\eta =\frac{2\pi \sqrt{Cv}{{\left( \frac{{{e}^{\lambda {{S}_{R}}}}-1}{\lambda } \right)}^{1/2}}{{{\hat{E}}}_{R}}-4\pi C\left( \frac{{{e}^{\lambda {{S}_{R}}}}-1}{\lambda } \right)-{{\pi }^{2}}{{{\hat{Q}}}^{2}}}{{{\left( \frac{{{e}^{\lambda {{S}_{R}}}}-1}{\lambda } \right)}^{2}}}\notag \\
-2\pi \sqrt{Cv}{{\left( \frac{{{e}^{\lambda {{S}_{R}}}}-1}{\lambda } \right)}^{-3/2}}\frac{2B\sqrt{\pi C}}{\sqrt{v}}\ln \left( \frac{1}{\sqrt{\pi }B}{{\left( \frac{{{e}^{\lambda {{S}_{R}}}}-1}{\lambda } \right)}^{1/2}} \right)-1.
\end{align}

In the event that $\hat{E}_R=\hat{E}_R^0$(being constant), when $\tilde{Q}$, $C$, $B$, and $\nu$ are held constant, according to Eq. (\ref{4.25}), the following equation is obtained:

\begin{align}\label{4.26}
\frac{\partial {{S}_{R}}}{\partial \eta }=-\frac{\frac{1}{2\pi \sqrt{Cv}}{{\left( \frac{{{e}^{\lambda {{S}_{R}}}}-1}{\lambda } \right)}^{3/2}}}{\frac{4\pi C\left( \frac{{{e}^{\lambda {{S}_{R}}}}-1}{\lambda } \right)-{{\pi }^{2}}{{{\hat{Q}}}^{2}}+3(1+\eta ){{\left( \frac{{{e}^{\lambda {{S}_{R}}}}-1}{\lambda } \right)}^{2}}}{4\pi \sqrt{Cv}{{\left( \frac{{{e}^{\lambda {{S}_{R}}}}-1}{\lambda } \right)}^{3/2}}}{{e}^{\lambda {{S}_{R}}}}+\frac{B\sqrt{\pi C}}{{{S}_{BH}}\sqrt{v}}\left( \frac{\lambda }{{{e}^{\lambda {{S}_{R}}}}-1} \right){{e}^{\lambda {{S}_{R}}}}}.
\end{align}
In the event that entropy assumes the R$\acute{e}$nyi entropy, according to Eqs. (\ref{4.22}), (\ref{4.24}), and (\ref{4.26}), the GP relation within the CFT framework is obtained:
\begin{align}\label{4.27}
\frac{\partial \hat{E}_{R}^{0}}{\partial \eta }=\frac{1}{2\pi \sqrt{Cv}}{{\left( \frac{{{e}^{\lambda {{S}_{R}}}}-1}{\lambda } \right)}^{3/2}}=\underset{{{{\hat{E}}}_{R}}\to \hat{M}_{R}^{0}}{\mathop{\lim }}\,-{{\hat{T}}_{R}}\left( \frac{\partial {{S}_{R}}}{\partial \eta } \right).
\end{align}

\subsection{Universality relation from Sharma-Mittal entropy}

Another significant form of non-extensive entropy is the Sharma-Mittal entropy, which generalizes both R$\acute{e}$nyi and Tsallis entropies, is given by \cite{241102427,250100955,2005217224,170309355},
\begin{align}\label{4.28}
{{S}_{SM}}&=\frac{1}{\alpha }\left( {{(1+\beta {{S}_{T}})}^{\tfrac{\alpha }{\beta }}}-1 \right),~~{{S}_{BH}}=\frac{{{(\alpha {{S}_{SM}}+1)}^{\tfrac{\beta }{\alpha }}}-1}{\beta } \notag \\
{{S}_{SM}}&=\frac{1}{\alpha }\left( {{(1+\beta {{S}_{T}})}^{\tfrac{\alpha }{\beta }}}-1 \right),~~d{{S}_{SM}}={{(1+\beta {{S}_{BH}})}^{\tfrac{\alpha }{\beta }-1}}d{{S}_{BH}},~~\frac{\partial {{S}_{BH}}}{\partial {{S}_{SM}}}={{(1+\alpha {{S}_{SM}})}^{1-\tfrac{\beta }{\alpha }}}.
\end{align}
In this context, $S_T$  represents the Tsallis entropy, defined in terms of the horizon area $A=4{\pi}{r_{+}^{2}}$, where $r_+$ denotes the radius of the black hole event horizon. The parameters $\alpha$ and $\beta$ are adjustable and must be constrained through observational data. Notaly, in the limit $\alpha{\rightarrow}0$, the Sharma-Mittal entropy reduces to the R$\acute{e}$nyi entropy. Similarly, when $\alpha=\beta$, it reduces to Tsallis entropy.

Substituting Eq. (\ref{4.28}) into (\ref{3.4}), the following equation is obtained:
\begin{align}\label{4.29}
{{\tilde{M}}_{SM}}=\frac{\pi C\frac{{{(\alpha {{S}_{SM}}+1)}^{\tfrac{\beta }{\alpha }}}-1}{\beta }+{{\pi }^{2}}{{{\tilde{Q}}}^{2}}+(1+\eta ){{\left( \frac{{{(\alpha {{S}_{SM}}+1)}^{\tfrac{\beta }{\alpha }}}-1}{\beta } \right)}^{2}}}{2{{\pi }^{3/2}}l\sqrt{C}{{\left( \frac{{{(\alpha {{S}_{SM}}+1)}^{\tfrac{\beta }{\alpha }}}-1}{\beta } \right)}^{1/2}}}+\frac{{\tilde{\gamma }}}{2{{l}^{2}}}\ln \left( \sqrt{\frac{C}{\pi }}\frac{l}{{\tilde{\gamma }}}{{\left( \frac{{{(\alpha {{S}_{SM}}+1)}^{\tfrac{\beta }{\alpha }}}-1}{\beta } \right)}^{1/2}} \right),
\end{align}
\begin{align}\label{4.30}
{{\tilde{T}}_{SM}}&={{\left( \frac{\partial {{{\tilde{M}}}_{SM}}}{\partial {{S}_{R}}} \right)}_{\tilde{Q},C,\tilde{\gamma }}}=\frac{\pi C\frac{{{(\alpha {{S}_{SM}}+1)}^{\tfrac{\beta }{\alpha }}}-1}{\beta }-{{\pi }^{2}}{{{\tilde{Q}}}^{2}}+3(1+\eta ){{\left( \frac{{{(\alpha {{S}_{SM}}+1)}^{\tfrac{\beta }{\alpha }}}-1}{\beta } \right)}^{2}}}{4{{\pi }^{3/2}}l\sqrt{C{{\left( \frac{{{(\alpha {{S}_{SM}}+1)}^{\tfrac{\beta }{\alpha }}}-1}{\beta } \right)}^{3/2}}}}{{(1+\alpha {{S}_{SM}})}^{1-\tfrac{\beta }{\alpha }}} \notag \\
&+\frac{{\tilde{\gamma }}}{4{{l}^{2}}}\left( \frac{\lambda }{{{e}^{\lambda {{S}_{R}}}}-1} \right){{(1+\alpha {{S}_{SM}})}^{1-\tfrac{\beta }{\alpha }}}.
\end{align}
Accordingly, the first law of thermodynamics (Eq. (\ref{3.1})) that is satisfied by the black hole when the modified entropy is the state parameter of the black hole thermodynamic system, is reformulated as follows:
\begin{align}\label{4.31}
d{{\tilde{M}}_{SM}}={{\tilde{T}}_{SM}}d{{S}_{SM}}+\tilde{\varphi }d\tilde{Q}+\mu dC+\tilde{\prod }d\tilde{\gamma }.
\end{align}
When $\tilde{Q}$, $C$, $\tilde{\gamma}$ and $l$ are held constant, according to Eq.(\ref{4.29}), the following equation is obtained:
\begin{align}\label{4.32}
\frac{\partial {{{\tilde{M}}}_{SM}}}{\partial \eta }=\frac{1}{2{{\pi }^{3/2}}l\sqrt{C}}{{\left( \frac{{{(\alpha {{S}_{SM}}+1)}^{\tfrac{\beta }{\alpha }}}-1}{\beta } \right)}^{3/2}},
\end{align}
and
\begin{align}\label{4.33}
\eta& =\frac{2{{\pi }^{3/2}}l\sqrt{C}{{\left( \frac{{{(\alpha {{S}_{SM}}+1)}^{\tfrac{\beta }{\alpha }}}-1}{\beta } \right)}^{1/2}}{{{\tilde{M}}}_{SM}}-\pi C\frac{{{(\alpha {{S}_{SM}}+1)}^{\tfrac{\beta }{\alpha }}}-1}{\beta }-{{\pi }^{2}}{{{\tilde{Q}}}^{2}}}{{{\left( \frac{{{(\alpha {{S}_{SM}}+1)}^{\tfrac{\beta }{\alpha }}}-1}{\beta } \right)}^{2}}}\notag \\
&-{{\pi }^{3/2}}\sqrt{C}{{\left( \frac{{{(\alpha {{S}_{SM}}+1)}^{\tfrac{\beta }{\alpha }}}-1}{\beta } \right)}^{-3/2}}\frac{{\tilde{\gamma }}}{l}\ln \left( \sqrt{\frac{C}{\pi }}\frac{l}{{\tilde{\gamma }}}{{\left( \frac{{{(\alpha {{S}_{SM}}+1)}^{\tfrac{\beta }{\alpha }}}-1}{\beta } \right)}^{1/2}} \right)-1.
\end{align}
In the event that ${{{\tilde{M}}}_{SM}}={{{\tilde{M}}}_{SM}^{0}}$(being constant), when $\tilde{Q}$, $C$, $\tilde{\gamma}$, and $l$ are held constant, according to Eqs. (\ref{4.30}) and (\ref{4.33}), the following equation is obtained:
\begin{align}\label{4.34}
{{\left( \frac{\partial {{S}_{SM}}}{\partial \eta } \right)}_{\tilde{M}_{SM}^{0}}}=-\frac{\frac{1}{2{{\pi }^{3/2}}l\sqrt{C}}{{\left( \frac{{{(\alpha {{S}_{SM}}+1)}^{\tfrac{\beta }{\alpha }}}-1}{\beta } \right)}^{3/2}}}{{{{\tilde{T}}}_{SM}}}.
\end{align}
In the event that entropy assumes the Sharma-Mittal R$\acute{e}$nyi entropy, according to Eqs. (\ref{4.32}) and (\ref{4.34}), the GP relation within the RPS framework is obtained:
\begin{align}\label{4.35}
\frac{\partial \tilde{M}_{SM}^{0}}{\partial \eta }=\frac{1}{2{{\pi }^{3/2}}l\sqrt{C}}{{\left( \frac{{{(\alpha {{S}_{SM}}+1)}^{\tfrac{\beta }{\alpha }}}-1}{\beta } \right)}^{3/2}}=\underset{{{{\tilde{M}}}_{SM}}\to \tilde{M}_{SM}^{0}}{\mathop{\lim }}\,-{{\tilde{T}}_{SM}}\left( \frac{\partial {{S}_{SM}}}{\partial \eta } \right)
\end{align}

Substituting Eq. (\ref{4.28}) into (\ref{3.11}), the following equation is obtained:
\begin{align}\label{4.36}
{{\hat{E}}_{SM}}&=\frac{4\pi C\frac{{{(\alpha {{S}_{SM}}+1)}^{\tfrac{\beta }{\alpha }}}-1}{\beta }+{{\pi }^{2}}{{{\hat{Q}}}^{2}}+(1+\eta ){{\left( \frac{{{(\alpha {{S}_{SM}}+1)}^{\tfrac{\beta }{\alpha }}}-1}{\beta } \right)}^{2}}}{2\pi \sqrt{Cv}{{\left( \frac{{{(\alpha {{S}_{SM}}+1)}^{\tfrac{\beta }{\alpha }}}-1}{\beta } \right)}^{1/2}}}\notag \\
&+\frac{2B\sqrt{\pi C}}{\sqrt{v}}\ln \left( \frac{1}{\sqrt{\pi }B}{{\left( \frac{{{(\alpha {{S}_{SM}}+1)}^{\tfrac{\beta }{\alpha }}}-1}{\beta } \right)}^{1/2}} \right),
\end{align}
\begin{align}\label{4.37}
{{\hat{T}}_{SM}}&={{\left( \frac{\partial {{{\hat{E}}}_{SM}}}{\partial {{S}_{SM}}} \right)}_{\hat{Q},C,B,v}}=\frac{4\pi C\frac{{{(\alpha {{S}_{SM}}+1)}^{\tfrac{\beta }{\alpha }}}-1}{\beta }-{{\pi }^{2}}{{{\hat{Q}}}^{2}}+3(1+\eta ){{\left( \frac{{{(\alpha {{S}_{SM}}+1)}^{\tfrac{\beta }{\alpha }}}-1}{\beta } \right)}^{2}}}{4\pi \sqrt{Cv}{{\left( \frac{{{(\alpha {{S}_{SM}}+1)}^{\tfrac{\beta }{\alpha }}}-1}{\beta } \right)}^{3/2}}}\notag \\ &+\frac{B\sqrt{\pi C}}{{{S}_{BH}}\sqrt{v}}\frac{\beta }{{{(\alpha {{S}_{SM}}+1)}^{\tfrac{\beta }{\alpha }}}-1}.
\end{align}
Accordingly, the first law of thermodynamics (Eq. (\ref{3.9})) that is satisfied by the black hole when the modified entropy is the state parameter of the black hole thermodynamic system, is reformulated as follows:
\begin{align}\label{4.38}
d{{\hat{E}}_{SM}}&={{\hat{T}}_{SM}}d{{S}_{SM}}+\hat{\phi }d\hat{Q}-pdv+\mu dC+\hat{\prod }d\hat{\gamma }.
\end{align}
When $\tilde{Q}$, $C$, $B$ and $\nu$ are held constant, according to Eq.(\ref{4.36}), the following equation is obtained:
\begin{align}\label{4.39}
\frac{\partial {{{\hat{E}}}_{SM}}}{\partial \eta }&=\frac{{{\left( \frac{{{(\alpha {{S}_{SM}}+1)}^{\tfrac{\beta }{\alpha }}}-1}{\beta } \right)}^{3/2}}}{2\pi \sqrt{Cv}},\notag \\
\end{align}
and
\begin{align}\label{4.40}
\eta& =\frac{{{{\hat{E}}}_{SM}}2\pi \sqrt{Cv}{{\left( \frac{{{(\alpha {{S}_{SM}}+1)}^{\tfrac{\beta }{\alpha }}}-1}{\beta } \right)}^{1/2}}-4\pi C\frac{{{(\alpha {{S}_{SM}}+1)}^{\tfrac{\beta }{\alpha }}}-1}{\beta }-{{\pi }^{2}}{{{\hat{Q}}}^{2}}}{{{\left( \frac{{{(\alpha {{S}_{SM}}+1)}^{\tfrac{\beta }{\alpha }}}-1}{\beta } \right)}^{2}}} \notag \\
&-4B{{\pi }^{3/2}}C{{\left( \frac{{{(\alpha {{S}_{SM}}+1)}^{\tfrac{\beta }{\alpha }}}-1}{\beta } \right)}^{-3/2}}\ln \left( \frac{1}{\sqrt{\pi }B}{{\left( \frac{{{(\alpha {{S}_{SM}}+1)}^{\tfrac{\beta }{\alpha }}}-1}{\beta } \right)}^{1/2}} \right)-1.
\end{align}
In the event that $\hat{E}_{SM}=\hat{E}_{SM}^0$(being constant), when $\tilde{Q}$, $C$, $B$ and $\nu$ are held constant, according to Eqs. (\ref{4.37}) and (\ref{4.39}), the following equation is obtained:
\begin{align}\label{4.41}
{{\left( \frac{\partial {{S}_{SM}}}{\partial \eta } \right)}_{{{{\hat{E}}}_{SM}}}}=-\frac{{{\left( \frac{{{(\alpha {{S}_{SM}}+1)}^{\tfrac{\beta }{\alpha }}}-1}{\beta } \right)}^{3/2}}}{2\pi \sqrt{Cv}{{{\hat{T}}}_{SM}}}.
\end{align}
In the event that entropy assumes the Sharma-Mittal R$\acute{e}$nyi entropy, according to Eqs. (\ref{4.3}), (\ref{4.40}), and (\ref{4.41}), the GP relation within the CFT framework is obtained:
\begin{align}\label{4.42}
\frac{\partial \hat{E}_{SM}^{0}}{\partial \eta }=\frac{1}{2\pi \sqrt{Cv}}{{\left( \frac{{{(\alpha {{S}_{SM}}+1)}^{\tfrac{\beta }{\alpha }}}-1}{\beta } \right)}^{3/2}}=\underset{{{{\hat{E}}}_{SM}}\to \hat{E}_{SM}^{0}}{\mathop{\lim }}\,-{{\hat{T}}_{SM}}\left( \frac{\partial {{S}_{SM}}}{\partial \eta } \right).
\end{align}
This result is consistent with findings derived from black hole models based on Bekenstein-Hawking entropy \cite{220503648,82112}, further substantiating the assertion that the Goon-Penco (GP) relation constitutes a universal thermodynamic expression valid across all types of black holes. The GP relation imposes no stringent constraints on the specific form of the black hole's state parameters, provided that the state variables of the system satisfy the first law of thermodynamics. In such circumstances, the GP relation is satisfied between the black hole's mass $M$, entropy $S$, temperature $T$, and perturbation parameter $\eta$.

\section{discussion}\label{five}

Recent progress in the field of black hole thermodynamics has led to the development of various corrective terms with the aim of refining the relationship between black hole entropy and the extremality bound. While this connection has traditionally been studied within the Bekenstein-Hawking entropy framework, the present work extends this scope by incorporating non-extensive generalizations of Gibbs entropy, thereby offering a more comprehensive thermodynamic description of black holes.

In the present study, a minor constant correction, denoted as $\eta$, and universal relations for Reissner-Nordstr$\ddot{o}$m Anti-de Sitter (AdS) black holes surrounded by perfect fluid dark matter (PFDM) were investigated. The results obtained demonstrates that even when the form of black hole entropy is generalized beyond the Bekenstein-Hawking case, the Goon-Penco (GP) relation remains invariant and unaffected by the specific form of the entropy. Subsequent analysis indicates that this relation is independent of not only of the entropy modification, but also of the selection of thermodynamic framework, encompassing both the RPS and CFT.

These findings indicate that the validity of the GP relation is contingent on the satisfaction of the first law of thermodynamics, a universal relation that is independent of the choice of state variables. Consequently, the GP relation is universal and robust against changes in the selection of black hole state parameters. Notwithstanding the application of modifications to the Bekenstein-Hawking entropy or the Hawking temperature, the GP relation remains unaltered, provided that the state variables of system continue to obey the first law of thermodynamic.

Furthermore, the proportional relationship between corrected mass and entropy provides a constructive pathway to towards elucidating the Weak Gravity Conjecture (WGC), a pivotal concept in quantum gravity that provides insights into the fundamental structure of physical theories.  Research endeavours of this nature significantly advance our understanding of fundamental aspects of theoretical physics.

\section*{Acknowledgments}
We would like to thank Prof. Ren Zhao for their indispensable discussions and comments. This work was supported by the Natural Science Foundation of China (Grant No. 12375050), Shanxi Provincial Natural Science Foundation of China (Grant No. 2025030221211241,~202203021221211)

\end{document}